\documentclass[12pt]{article}
\usepackage{graphicx}
\usepackage{cite}
\usepackage{hyperref}
\usepackage[belowskip=-11pt,aboveskip=0pt]{caption}

\def \confproc{Proceedings of CKM 2012, the 7th International Workshop on the
CKM Unitarity Triangle, University of Cincinnati, USA, 28 September - 2 October
2012}

\def \bea{\begin{eqnarray}}
\def \eea{\end{eqnarray}}
\def \beq{\begin{equation}}
\def \eeq{\end{equation}}
\def \nn{\nonumber}
\def \ok{\overline{K}^0}
\def \({\left(}
\def \){\right)}
\def \s{\sqrt{2}}
\def \ca{{\cal A}}
\def \cao{\overline{\ca}}

\def \KK{K^+K^-}
\def \pp{\pi^+\pi^-}
\def \i{\it i}
\def\Title#1{\begin{center} {\Large {\bf #1} } \end{center}}

\begin{document}
\rightline{UdeM-GPP-TH-12-218}
\rightline{February 2013}

\Title{Direct CP Violation in Nonleptonic Charm Decays\footnote{\confproc}}
\smallskip


\begin{raggedright}

{\it Bhubanjyoti Bhattacharya \\ 
Physique des Particules, Universit\'e
de Montr\'eal,\\
\it C.P. 6128, succ.\ centre-ville, Montr\'eal, QC,
Canada H3C 3J7}
\end{raggedright}

\smallskip
\smallskip


\begin{quote}
LHCb, CDF and most recently Belle have observed a large difference between
direct CP asymmetries in $D^0\to\KK$ and $D^0\to\pp$. We use the measured
value of $\Delta A_{CP}$ to constrain the magnitude and strong phase of a
QCD-enhanced standard model penguin and predict its effects on direct CP
asymmetry in $D^0\to\pi^0\pi^0$ and $D^+\to K^+\ok$.
\end{quote}

\smallskip

The LHCb and CDF Collaborations have reported evidence for close to a percent%
-level CP violation \cite{Aaij:2011in} in the difference between direct CP
asymmetries ($\Delta A_{CP}$) in $D^0\to\KK$ and $D^0\to\pp$. Their result has
very recently been corroborated by the Belle Collaboration \cite{Ko:2012}. The
conventional standard model (SM) prediction for $\Delta A_{CP}$, however, is at
least an order of magnitude smaller due to Cabibbo-Kobayashi-Maskawa (CKM)
suppression. The deviation of the measured $\Delta A_{CP}$ from its SM prediction
has naturally led to the discussion of whether this is a new signature of physics
beyond the SM. However, the closeness of the charm-quark mass ($m_c$) to the scale
at which perturbative QCD breaks down ($\Lambda_{QCD}$) presents an opportunity
to rescue SM physics. The hope is that non-perturbative QCD contributions can
generate enough enhancement to compensate for the large CKM suppression.

Although it is quite early to conclude whether or not this is new physics, a useful
exercise is to constrain the SM parameters using these measurements and to predict
their effects on other $D$-meson decays where CP-asymmetries are yet to be measured.
In this talk, we present our analysis of $D$ meson decay rates and CP asymmetries
based on flavor-SU(3) symmetry. We assume that an enhanced SM $c\to b\to u$ penguin
is responsible for the large measured $\Delta A_{CP}$ and we quantify its effects
on direct CP asymmetry in $D^0\to\pi^0\pi^0$ and $D^+\to\ok K^+$. The analysis
presented in this talk closely follows that in Ref.\ \cite{Bhattacharya:2012ah},
with the addition of new updates from a Belle preliminary analysis \cite{Ko:2012}.

$D$ meson decays to two pseudoscalar mesons were studied in Ref.\
\cite{Bhattacharya:2008ss}. $D$ decays are of three kinds: a) Cabibbo-favored (CF) (%
proportional to $V^*_{cs}V_{ud}$), b) singly-Cabibbo-suppressed (SCS) (proportional
to $V^*_{cd}V_{ud}$ or $V^*_{cs}V_{us}$) and c) doubly-Cabibbo-suppressed (DCS) (%
proportional to $V^*_{cd}V_{us}$). Since the CF and DCS transitions involve
distinct quark flavors, these amplitudes don't get penguin contributions. In Ref.\
\cite{Bhattacharya:2008ss} it was shown that flavor-SU(3) symmetry using tree-level
topologies $T$, $C$, $E$ and $A$ is able to fit the measured CF decay amplitudes with
good precision. Under flavor-SU(3) symmetry the SCS decay amplitudes for $D^0\to\pp$
and $D^0\to\KK$ are expected to obey:
\beq
\ca(D^0\to\KK) = - \ca(D^0\to\pp) = \lambda (T + E)~,
\eeq
where $\lambda = \tan\theta_{\rm Cabibbo} = 0.2317$. Also the amplitude for $D^0\to
K^0\ok$ is expected to vanish. In practice, however, the measured amplitude for $D^0
\to K^0\ok$ is found to be 239$\pm$14 eV while that for $D^0 \to\pp(\KK)$ is found to
be smaller (larger) than the corresponding flavor-SU(3) prediction.

Factorizable flavor-SU(3) breaking effects in the amplitude $T$ can be calculated in
terms of meson masses ($m_K$, $m_\pi$) and decay constants ($f_K$, $f_\pi$):
\bea
T_\pi &=& T\,\cdot\,\frac{|f_{+(D^0\to\pi^-)}(m^2_\pi)|}{|f_{+(D^0\to K^-)
(m^2_\pi)|}} \,\cdot\, \frac{m^2_D - m^2_\pi}{m^2_D - m^2_K}~, \\
T_K   &=& T\,\cdot\,\frac{|f_{+(D^0\to  K^-)}(m^2_K)|}{|f_{+(D^0\to K^-)}
(m^2_\pi)|} \,\cdot\,\frac{f_K}{f_\pi}.
\eea
These corrections diminish the discrepancy between theory and measured amplitudes, but
they are not enough. Furthermore, factorizable corrections do not affect the amplitude for
$D^0\to K^0\ok$. We recall that the SCS decay amplitudes can get contributions from
penguin topologies which are ordinarily suppressed by the Glashow-Iliopulos-Maiani (GIM)
mechanism, but which may be nonzero if flavor SU(3) is broken. Once these penguins are
included we have additional parameters that may be extracted from the data as shown in Ref.\
\cite{Bhattacharya:2012ah}. In Table \ref{tab:amps} we present the flavor-topology
representations for the SCS $D$ decays to pions and kaons, alongside the theory fit and
measured amplitudes. We have also listed the overall strong phase ($\phi^f_T$) for each
process.
\begin{table}[t!]
\caption{Representations and comparison of experimental and fit
amplitudes for SCS decays of charmed mesons to two pseudoscalar mesons.
Also listed is the overall strong phase ($\phi^f_T$) for each process
\cite{Bhattacharya:2012ah}.
\label{tab:amps}}
\centering
\begin{tabular}{|l c c c c|} \hline \hline
Decay & Amplitude      & \multicolumn{2}{c}{$|\ca|$ ($10^{-7}$ GeV)} & $\phi^f_T$\\
\cline{3-4}
Mode & representation & Experiment & Theory & degrees\\ \hline
$D^0\to\pi^+\pi^-$ &$-\lambda\,(T_\pi + E) + (P + PA)$  &4.70$\pm$0.08&4.70&--158.5\\
$D^0\to  K^+  K^-$ &~$\lambda\,(T_K + E) + (P + PA)$ &8.49$\pm$0.10&8.48&32.5\\
$D^0\to\pi^0\pi^0$ &$-\lambda\,(C - E)/\s - (P + PA)/\s$&3.51$\pm$0.11&3.51&60.0\\ \hline
$D^+\to\pi^+\pi^0$ &$-\lambda\,(T_\pi + C)/\s$ &2.66$\pm$0.07&2.26&126.3\\ \hline
$D^0\to K^0\ok$    &$-(P + PA) + P$                 &2.39$\pm$0.14&2.37&--145.6\\
$D^+\to K^+\ok$    &~$\lambda\,(T_K - A_{D^+}) + P$    &6.55$\pm$0.12&6.87&--4.2\\
$D^+_s\to\pi^+ K^0$&$-\lambda\,(T_\pi - A) + P$        &5.94$\pm$0.32&7.96&174.3\\
$D^+_s\to\pi^0 K^+$&$-\lambda\,(C + A)/\s - P/\s$      &2.94$\pm$0.55&4.44&16.4\\
\hline \hline
\end{tabular}
\end{table}

The tiny weak-phase difference between the tree-level amplitudes and the GIM-suppressed penguins 
$P$ and $PA$ is insufficient to produce a $\Delta A_{CP}$ as large as measured by LHCb, CDF and
Belle. The measured $\Delta A_{CP}$ calls for an enhanced $c\to b\to u$ penguin denoted by
$P_b$ \footnote{Note that both $P, PA$ and $P_b$ require the same enhancement as explained
in detail in Ref.\ \cite{Bhattacharya:2012ah}}. Let us parameterize the CP-conserving part
of an SCS amplitude as $T_f = |T_f| e^{\i\phi^f_T}$ and the CP-violating penguin part as
$P_b = p e^{\i(\delta - \gamma)}$, where $\delta$ and $\gamma$ are respectively its strong and
weak phases. We use $\gamma = 76^\circ$. The amplitude for a general SCS $D$ decay to a final
state $f$ is
\beq
\ca(D\to f)~~=~~T_f + P_b~~=~~|T_f|e^{\i\phi^f_T}\left(1 + \frac{p}{|T_f|}e^{\i(\delta
- \phi^f_T- \gamma)}\right)~. \label{eqn:A}
\eeq
The amplitude for the CP-conjugate process $\cao(\bar D\to\bar f)$ can be obtained by changing
the sign of the weak phase $\gamma$ in Eq.\ (\ref{eqn:A}). The direct CP asymmetry in $D\to f$
can then be expressed as:
\bea
A_{CP}(f) &=& \frac{|\ca|^2 - |\cao|^2}{|\ca|^2 + |\cao|^2}
~~=~~\frac{2(p/|T_f|)\sin\gamma\sin(\delta - \phi^f_T)}{1 + (p/|T_f|)^2
+ 2(p/|T_f|)\cos\gamma\cos(\delta - \phi^f_T)} \nn \\
&\approx& 2(p/|T_f|)\sin\gamma\sin(\delta - \phi^f_T)~,\label{eqn:CPV}
\eea

\begin{figure}[t!]
\centering
             \includegraphics[width=0.6\textwidth]{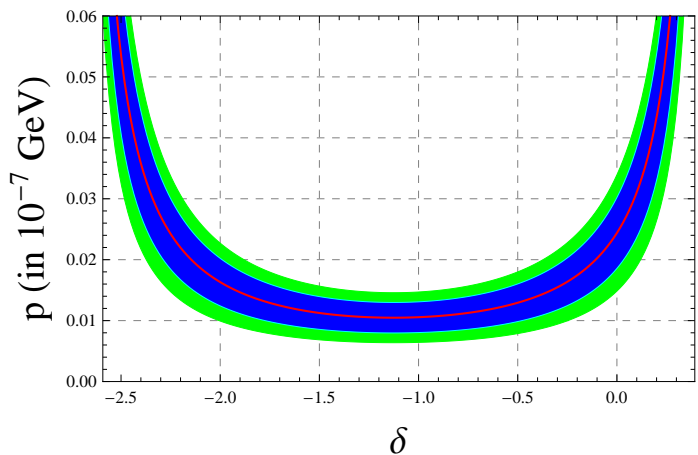}
\caption{$p$ and $\delta$ allowed by the measured range of $\Delta A_{CP}$. The (red)
line represents the central value, while inner (blue) and outer (green) bands
respectively represent 68\% (1$\sigma$) and 90\% (1.64$\sigma$) confidence-level
regions based on error in $\Delta A_{CP}$. \label{fig:pb}}
\end{figure}

The measured value of $\Delta A_{CP}$ from LHCb, and CDF \cite{Aaij:2011in} is
\beq
\Delta A_{CP}~~=~~A_{CP}(\KK) - A_{CP}(\pp)~~=~~(-0.67\pm0.16)\%
\eeq
We constrain $p$ as a function of $\delta$ using Eq.\ (\ref{eqn:CPV}) along with the
measured value of $\Delta A_{CP}$. In addition, we use the Belle preliminary results
\cite{Ko:2012} as bounds for the individual CP asymmetries in $D^0\to\KK$ and $D^0\to\pp$:
\beq
A_{CP}(D^0\to\KK) = (-0.23\pm0.15)\%~,~~~~~ A_{CP}(D^0\to\pp) = (0.27 \pm 0.20)\%~,
\eeq
from which we calculate the following 90\% confidence-level limits:
\bea
-0.48\%\leq A_{CP}(\KK)\leq0.02\%~,~~~~~-0.06\%\leq A_{CP}(\pp)\leq0.60\%~.
\eea
These numbers put a bound on the range of allowed values of $\delta$ to $-2.59\leq\delta
\leq0.39$. (The CP asymmetries are approximately invariant under the joint transformations
$\phi^f_T\to -\phi^f_T$ and $\delta\to\pi - \delta$ as long as $p/|T_f|$ is small compared
to unity.) In Fig.\ \ref{fig:pb} we plot $p$ as a function of $\delta$ within this allowed
range. In this plot $p$ stays close to $10^{-9}$ GeV for a large range of $\delta$.

\begin{figure}[t!]
\centering
\includegraphics[width=0.46\textwidth]{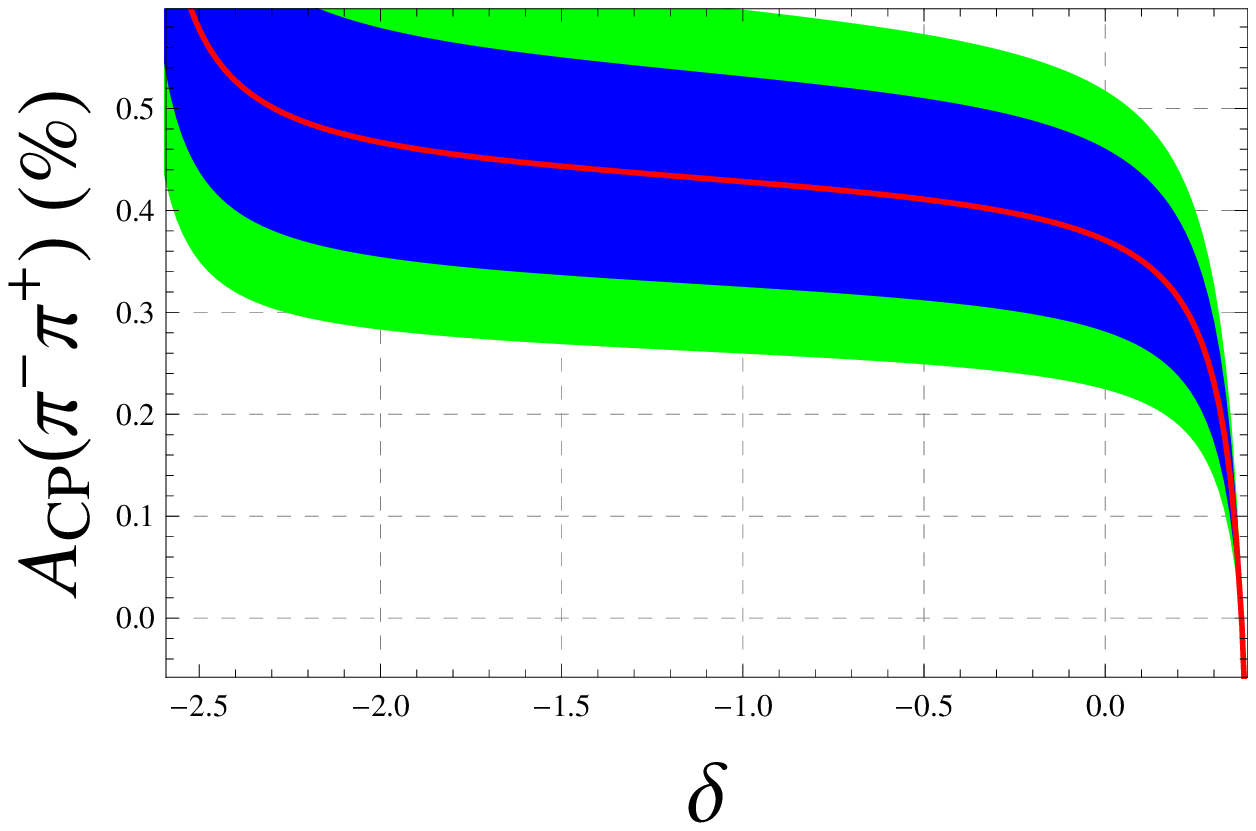} \hspace{0.15in}
\includegraphics[width=0.465\textwidth]{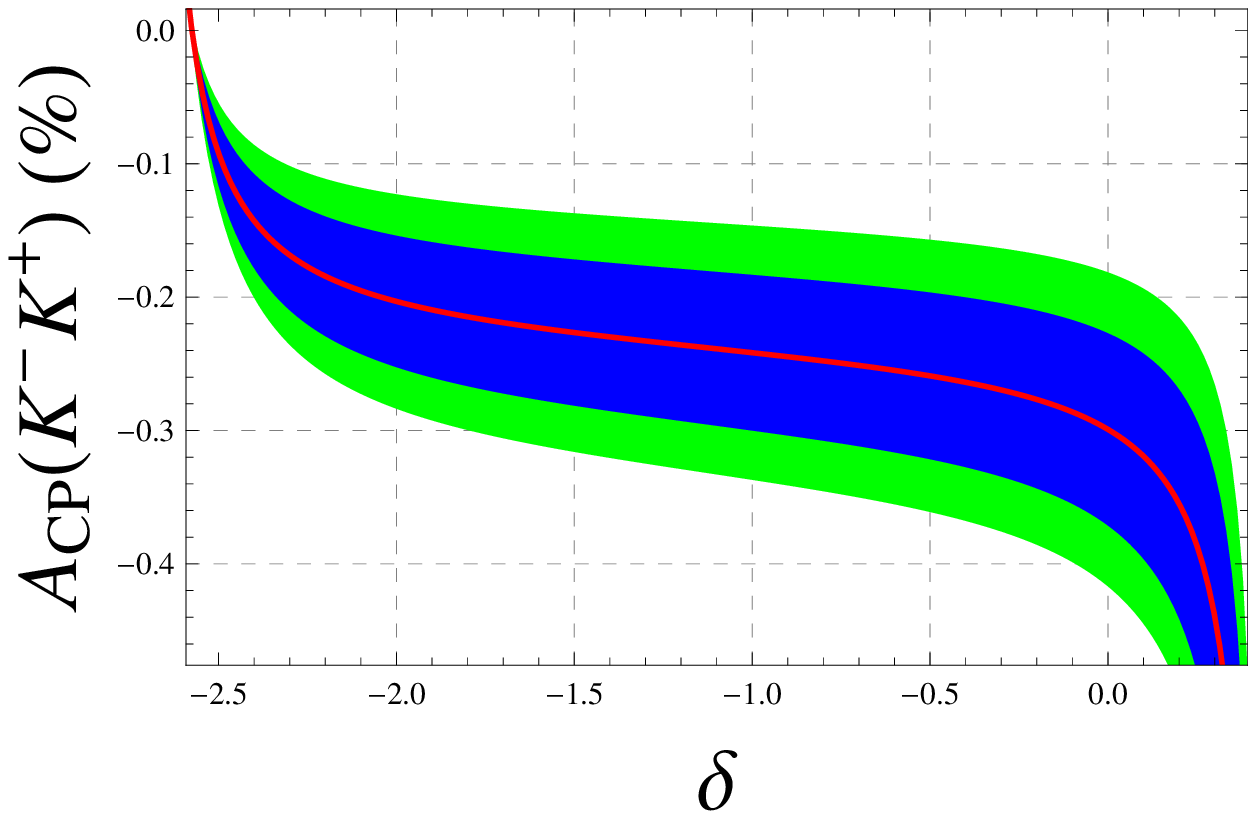}
\includegraphics[width=0.455\textwidth]{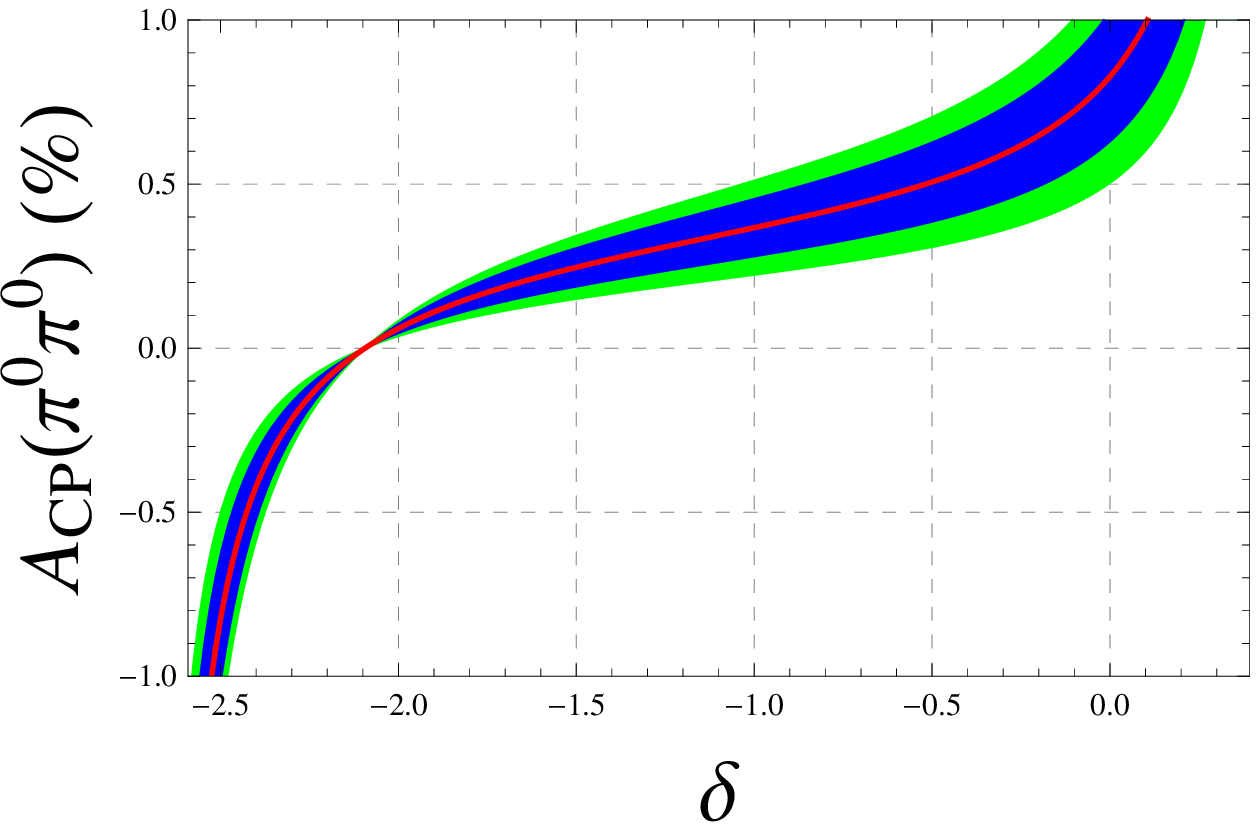} \hspace{0.15in}
\includegraphics[width=0.47\textwidth]{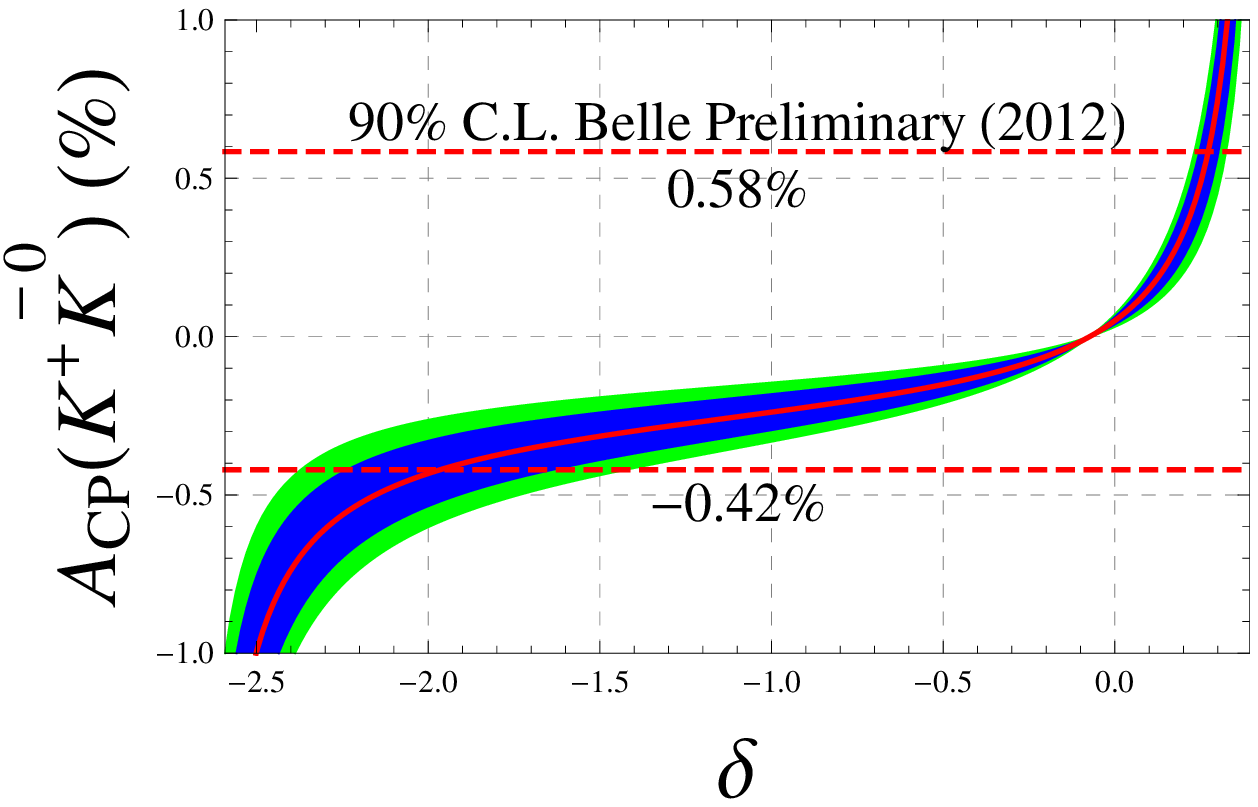}
\caption{Direct CP asymmetry in various $D^0$ and $D^+$ decay modes plotted as a
function of $\delta$. The red dashed lines in the lower-right-hand plot shows 90\%
(1.64$\sigma$) confidence-level preliminary bounds from Belle, 2012 \cite{Ko:2012}.
The solid (red) lines represent the central values, while inner (blue) and outer
(green) bands respectively represent 68\% (1$\sigma$) and 90\% (1.64$\sigma$)
confidence-level regions based on error in $\Delta A_{CP}$. \label{fig:ACP}}
\end{figure}

We now use the $p-\delta$ constraint to find its effects on the individual CP asymmetries
$D^0\to\pp$ and $D^0\to\KK$ as well as predict the CP asymmetries in $D^0\to\pi^0\pi^0$ and
$D^+\to K^+\ok$. In Fig.\ \ref{fig:ACP} we plot the individual CP asymmetries as a function
of $\delta$. The plots show that the direct CP asymmetries are correlated and that for a
large range of $\delta$, $A_{CP}$ is positive (negative) for $D^0\to\pp$ and $D^0\to\pi^0\pi
^0$ ($D^0\to\KK$ and $D^+\to K^+\ok$). Preliminary results from Belle, 2012 \cite{Ko:2012}
indicate $A_{CP}(D^+\to K^+\ok) = (0.082\pm0.275\pm0.135)\%$. The corresponding 90\%
confidence-level bounds are shown as red dashed lines in the lower right panel plot of Fig.\
\ref{fig:ACP}. A large range of $\delta$ is still allowed by the measured CP asymmetries.
In this talk, we have presented a model based on flavor topologies that accounts well for
the measured decay rates in CF $D$ decays. SU(3) breaking in SCS decays is taken care of
by introducing the ordinarily GIM-suppressed penguin as a parameter. The LHCb, and CDF
observed large difference in CP asymmetries between $D^0\to\KK$ and $D^0\to\pp$ is used to
constrain the magnitude and strong phase of a $c\to b\to u$ penguin, which is then used to
predict CP asymmetries in $D^0\to\pi^0\pi^0$ and $D^+\to K^+\ok$. In our model we have
neglected a CP-violating annihilation penguin, which may give rise to $A_{CP}$ in $D^0\to
K^0\ok$. Under isospin symmetry the amplitude for $D^+\to\pi^+\pi^0$ doesn't involve
penguins, and hence the associated direct CP asymmetry vanishes. Future measurements may
make it possible to apply this analysis to $D^+_s$ decays as well as $D$ decays to a
pseudoscalar and a vector.

\bigskip
This work was financially supported by the NSERC of Canada. Thanks to Jon Rosner
and Michael Gronau for a fruitful collaboration. Thanks to Alexey Petrov, Gil Paz
and Anze Zupanc for useful conversations. Thanks also to the conference organizers
for a wonderful stay in Cincinnati.

%
%
%

\end{document}